\documentclass[aps,prc,twocolumn, superscriptaddress,showkeys,nofootinbib]{revtex4-1}  

\usepackage[utf8]{inputenc}

\usepackage{graphicx,color}
\usepackage{amsmath,amssymb,amsfonts}
\usepackage{hyperref}

\usepackage{xspace}	

\newcommand{\roots} {\mbox{$\sqrt{\textit{s}_{NN}}$}\xspace}

\def  \pT         {\mbox{$p_{T}$}\xspace}

\def  \RuRu       {\mbox{$\rm Ru+Ru$ }\xspace}
\def  \ZrZr       {\mbox{$\rm Zr+Zr$ }\xspace}

\begin{document}
\title{
Exploring Rapidity-Even Dipolar Flow in Isobaric Collisions at RHIC
}
\medskip
\author{Niseem~Magdy} 
\email{niseemm@gmail.com}
\affiliation{Center for Frontiers in Nuclear Science at the State University of New York, Stony Brook, NY 11794, USA}
\affiliation{Department of Chemistry, the State University of New York, Stony Brook, New York 11794, USA}
\author{Roy~Lacey} 
\affiliation{Department of Chemistry, the State University of New York, Stony Brook, New York 11794, USA}
\affiliation{Department of Physics, Stony Brook University, Stony Brook, NY 11794}

\begin{abstract}
Employing the AMPT transport model, we investigate the response of the rapidity-even dipolar flow ($\mathrm{v^{even}_{1}}$) and its associated Global Momentum Conservation (GMC) parameter $\mathrm{K}$ to structural disparities within $^{96}$Ru and $^{96}$Zr nuclei. We analyze $\rm Ru+Ru$ and $\rm Zr+Zr$ collisions at a center-of-mass energy of $\sqrt{\textit{s}_{NN}}$ = 200~GeV. Our analysis demonstrates that the eccentricity $\varepsilon_1$, $\mathrm{v^{even}_{1}}$ and $\mathrm{K}$ exhibit subtle yet discernible sensitivity to the input nuclear structure distinctions between $^{96}$Ru and $^{96}$Zr isobars. This observation suggests that measuring $v^{\rm even}_{1}$ and the GMC parameter in these isobaric collisions could serve as a means to fine-tune the comprehension of their nuclear structure disparities and offer insights to enhance the initial condition assumptions of theoretical models.
\end{abstract}
\maketitle

High-energy nuclear collisions at the Relativistic Heavy Ion Collider (RHIC) and the Large Hadron Collider (LHC) can produce a strongly coupled plasma of quarks and gluons (QGP). Full characterization of this hot and dense matter is central to present-day high-energy physics research.  Such characterization requires detailed knowledge of the initial state in Heavy Ion Collisions (HIC) ~\cite{Danielewicz:1998vz, Ackermann:2000tr, Adcox:2002ms, Heinz:2001xi, Hirano:2005xf, Huovinen:2001cy, Hirano:2002ds, Romatschke:2007mq, Luzum:2011mm,
Song:2010mg,Qian:2016fpi,Schenke:2011tv,Teaney:2012ke,Gardim:2012yp,Lacey:2013eia,Magdy:2022ize}.
Numerous investigations have underscored the significance of anisotropic flow measurements, mainly focusing on n{th}-order flow harmonics ($v_n, \;_{n\ge 2}$), as a strategic avenue for studying both the initial conditions and transport properties of the QGP such as the specific viscosity $\eta/s$ or the ratio of viscosity to entropy density \cite{Teaney:2003kp, Lacey:2006pn,Song:2011qa, Alver:2010gr, Schenke:2010rr,Shen:2010uy, Niemi:2012ry, Qin:2010pf, Magdy:2023owx}.

Indeed, a wealth of research endeavors has yielded a treasure trove of insights into the initial conditions shaping HIC events and the properties of the QGP produced in them. These insights have been garnered through comprehensive studies of dipolar ($v_1$), elliptic ($v_2$) and triangular ($v_3$) flow harmonics, encompassing not only their magnitudes, statistical variances and fluctuations but also the intricate interplay between them~\cite{STAR:2018fpo, ALICE:2016kpq, Adamczyk:2017hdl, Qiu:2011iv, Adare:2011tg, Aad:2014fla, Aad:2015lwa, Magdy:2018itt, Alver:2008zza, Alver:2010rt, Ollitrault:2009ie, Magdy:2022jai, STAR:2022gki, Adamczyk:2016gfs, STAR:2015rxv, Magdy:2018itt, Adamczyk:2015obl, Adamczyk:2016exq, Adam:2019woz, Song:2010mg, Niemi:2012aj, Gardim:2014tya, Fu:2015wba, Holopainen:2010gz, Qin:2010pf, Qiu:2011iv, Gale:2012rq, Liu:2018hjh, Magdy:2022ize}.

Recent investigations have delved into the intriguing realm of exploiting collisions to constrain the structure of the colliding nuclei, specifically focusing on the ratios $\frac{v^{\rm Ru+Ru}_2}{v^{\rm Zr+Zr}_2}$ and $\frac{v^{\rm Ru+Ru}_3}{v^{\rm Zr+Zr}_3}$ obtained from collisions involving $^{96}$Ru+$^{96}$Ru and $^{96}$Zr+$^{96}$Zr~\cite{STAR:2021mii, Li:2018oec, Li:2019kkh, Xu:2021vpn, Xu:2021qjw, Xu:2021uar, Zhao:2022uhl, Xu:2022ikx, Wang:2023yis, Dimri:2023wup, Liu:2022xlm, Nie:2022gbg, Jia:2022qgl, Jia:2022qrq, Zhang:2022fou, Jia:2022iji, Liu:2022kvz}. This avenue has been explored as a means to impose constraints on the disparities in the initial-state deformation and the neutron skin of these two isobars.
The motivation behind these investigations arises from the notion that, in these isobaric collisions, there is an expectation that the viscosity-related attenuation dependent on final-state multiplicity should exhibit similarities~\cite{STAR:2019zaf}. However, subtle variations in their inherent shapes and neutron distributions can lead to distinct differences in the initial-state eccentricities~\cite{Jia:2021tzt, Giacalone:2021udy}, thereby giving rise to noticeable disparities in the magnitudes of $v_2$ and $v_3$ for the respective isobaric pairs.

{\color{black}The component of directed flow that is even in rapidity ($v^{\rm even}_1$), also known as dipolar flow, arises from the interplay of initial-state fluctuations and a hydrodynamic-like expansion process \cite{Luzum:2010fb,Teaney:2010vd,Gardim:2011qn,Magdy:2018whk,STAR:2018gji}. As a result, its magnitude is influenced by both the initial-state eccentricity ($\varepsilon_1$) and viscous effects that are dependent on multiplicity ($\mathrm{N_{ch}}$) and $\eta/s$. However, different studies suggest that $v^{\rm even}_1$ is less sensitive to viscous effects compared to higher-order harmonics ($n \geq 2$) \cite{Retinskaya:2012ky,STAR:2018gji}. Given these considerations, it is valuable to explore whether the subtle variations in intrinsic deformations and neutron skin thickness among the isobars will lead to characteristic distinctions in the behavior of $v^{\rm even}_1$ between them. Thus, we pose the question: \textit{To what extent will the difference in the initial state between Ru+Ru and Zr+Zr will impact the $v^{\rm even}_1$ observable?
}
}

In this investigation, we employ the multiphase transport model AMPT to analyze the behavior of both $\varepsilon_{1}$ and $v^{\rm even}_1$ with respect to collision centrality in scenarios involving $^{96}$Ru+$^{96}$Ru and $^{96}$Zr+$^{96}$Zr collisions at \roots = 200~GeV. These analyses take into account the distinct intrinsic deformations and neutron skin disparities assumed to be inherent to these isobaric species.

The AMPT model~\cite{Lin:2004en} is a comprehensive simulation framework widely utilized to delve into the complexities of relativistic heavy-ion collisions~\cite{Lin:2004en, Ma:2016fve, Haque:2019vgi, Bhaduri:2010wi, Nasim:2010hw, Xu:2010du, Magdy:2020bhd, Guo:2019joy}. In this particular study, we generate AMPT events that incorporate the string-melting option. Within the AMPT framework, the Glauber model is employed to define the shape and radial distribution of the colliding nuclei, employing a deformed Woods-Saxon distribution~\cite{Hagino:2006fj}:
\begin{eqnarray}\label{Eq:Rthita}
\rho(r, \theta, \phi) &\propto& \frac{1}{1+e^{(r-R(\theta, \phi))/a_0}}, \\
R(\theta, \phi) &=& R_0(1+\beta_2Y^0_2( \theta, \phi)+\beta_3Y^0_3( \theta, \phi)).
\end{eqnarray}
Here, the nuclear surface $R(\theta, \phi)$ incorporates quadrupole and octupole deformations regulated by $\beta_2$ and $\beta_3$, respectively, and the parameters $R_0$ and $a_0$ represent the half-height radius and nuclear skin thickness. In collisions, the projectile and target nuclei are rotated event-by-event randomly along the polar and azimuthal directions.
The Woods-Saxon parameter sets for $^{96}$Ru and $^{96}$Zr employed in our analysis are summarized in  Table~\ref{tab:1}~\cite{Moller:1993ed}.
\begin{table}[h!]
\begin{center}
 \begin{tabular}{|c|c|c|c|c|}
 \hline 
  AMPT-set      &   $R_{0}$ &  $a_{0}$  & $\beta_{2}$  &  $\beta_{3}$   \\
  \hline
  $^{96}$Ru+$^{96}$Ru (Case-1)    &    5.09   &  0.46   &  0.162       &   0.00           \\
  \hline
  $^{96}$Zr+$^{96}$Zr (Case-2)    &    5.02   &  0.52   &  0.06        &   0.20           \\
 \hline
  $^{96}$Zr+$^{96}$Zr (Case-3)    &    5.09   &  0.46   &  0.06        &   0.20           \\
 \hline
  $^{96}$Zr+$^{96}$Zr (Case-4)    &    5.09   &  0.46   &  0.06        &   0.00           \\
 \hline
\end{tabular} 
\caption{ Summary of the Woods-Saxon parameters for $^{96}$Ru and $^{96}$Zr employed in the AMPT studies.}
\label{tab:1}
\end{center}
\end{table}

The dipole asymmetry or eccentricity $\varepsilon_{1}$ of the collision systems, was obtained as a function of centrality for the respective cases outlined in Table~\ref{tab:1}, as~\cite{Luzum:2010fb, Teaney:2010vd, Gardim:2011qn}:
\begin{eqnarray}
\label{eq:ecc1}
\varepsilon_{1} &\equiv& - \dfrac{\langle r^{3} \cos(\phi - \psi_{1,3}) \rangle}{\langle r^{3} \rangle}
\end{eqnarray}
\begin{eqnarray}
\label{eq:psi1}
\psi_{1,3} &\equiv& \text{atan2}\left(\langle r^{3} \sin\phi \rangle, \langle r^{3} \cos\phi \rangle\right) + \pi
\end{eqnarray}
where the averaging is performed over the initial energy density. 

Within the same AMPT framework, the HIJING model is employed to generate hadrons, which are then transformed into their constituent quarks and anti-quarks. The subsequent space-time evolution of these particles is determined using the ZPC Parton cascade model~\cite{Zhang:1997ej}, which incorporates the parton-scattering cross-section:
\begin{eqnarray} \label{eq:21}
\sigma_{pp} &=& \dfrac{9 \pi \alpha^{2}_{s}}{2 \mu^{2}},
\end{eqnarray}
where $\alpha_{s}$ denotes the QCD coupling constant and $\mu$ is the screening mass in the partonic matter; these parameters collectively shape the expansion dynamics of the collision systems~\cite{Zhang:1997ej}.  Collisions were  simulated for a fixed value of {\color{black}$\sigma_{pp}$ = 2.8~(mb)}~\cite{Xu:2011fi,Nasim:2016rfv}. Thus, the AMPT model amalgamates several contributing factors: (i) the initial-state eccentricity, (ii) the initial Parton-production stage governed by the HIJING model~\cite{Wang:1991hta, Gyulassy:1994ew}, (iii) a Parton-scattering stage, (iv) the process of hadronization through coalescence, and finally, (v) a phase that encompasses interactions among the generated hadrons~\cite{Li:1995pra}.

The coefficients $v^{\rm even}_1$ were determined by studying the transverse momentum ($p_T$) and collision centrality dependencies using a two-particle correlation technique~\cite{Retinskaya:2012ky, ATLAS:2012at}. The analysis focused on charged hadrons within a transverse momentum range of $0.2$ $<$ $p_T$ $<$ $3.0$ GeV/c and a pseudorapidity acceptance of $|\eta|$ $<$ $1.0$. Additionally, simulated events were categorized into various collision centrality classes based on the collision's multiplicity.

Initially, the correlation function method~\cite{Lacey:2005qq} was employed to construct the two-particle $\Delta\phi$ correlations,
\begin{eqnarray}\label{corr_func}
C_r(\Delta\phi, \Delta\eta) = \frac{(dN/d\Delta\phi)_{\text{same}}}{(dN/d\Delta\phi)_{\text{mixed}}},
\end{eqnarray}
where $\Delta\eta = \eta_{a} - \eta_{b}$ represents the pseudorapidity difference between particles $a$ and $b$. The terms $(dN/d\Delta\phi)_{\text{same}}$ and $(dN/d\Delta\phi)_{\text{mixed}}$ denote the normalized azimuthal distribution for particle pairs originating from the same and mixed events, respectively. To mitigate short-range non-flow correlations~\cite{Lacey:2005qq, Borghini:2000cm, Luzum:2010fb, Retinskaya:2012ky, ATLAS:2012at}, a condition of $|\Delta\eta| > 0.7$ was applied to the pairs, which helps reduce contributions from resonance decays, Bose-Einstein correlations, and jets. However, it's important to note that long-range non-flow correlations~\cite{Retinskaya:2012ky, ATLAS:2012at} (such as those arising from jets and global momentum conservation (GMC)) still persist.

Subsequently, the values of $\mathrm{v_{11}}$ were deduced from the correlation functions (Eq.\ref{corr_func}) as:
\begin{eqnarray}\label{vn}
 \mathrm{v_{11}} &=& \frac{\sum_{\Delta\phi} C_r(\Delta\phi)\cos(n \Delta\phi)}{\sum_{\Delta\phi}~C_r(\Delta\phi)},
\end{eqnarray}
and simultaneously fitted as a function of $p_{T}^{\text{a}}$ and $p_{T}^{\text{b}}$ to extract the $\mathrm{v^{even}_{1}}$: 
\begin{equation}\label{corrv1}
\mathrm{v_{11}(\pT^{a},\pT^{b})  = v^{even}_{1}(\pT^{a})v^{even}_{1}(\pT^{b}) - K \pT^{a}\pT^{b}}.
\end{equation} 
Here the parameter $\mathrm{K}$ takes into account the long-range non-flow from GMC~\cite{Borghini:2000cm,Borghini:2002mv,Retinskaya:2012ky,ATLAS:2012at}. It is anticipated to be correlated with the average multiplicity $\mathrm{\langle N_{ch} \rangle}$ and the variance of the transverse momentum across the full phase space (i.e., $\mathrm{K \propto 1/(\langle N_{ch} \rangle \langle p_{T}^{2}\rangle)}$).

\begin{figure}[!h] 
\includegraphics[width=1.0  \linewidth, angle=-0,keepaspectratio=true,clip=true]{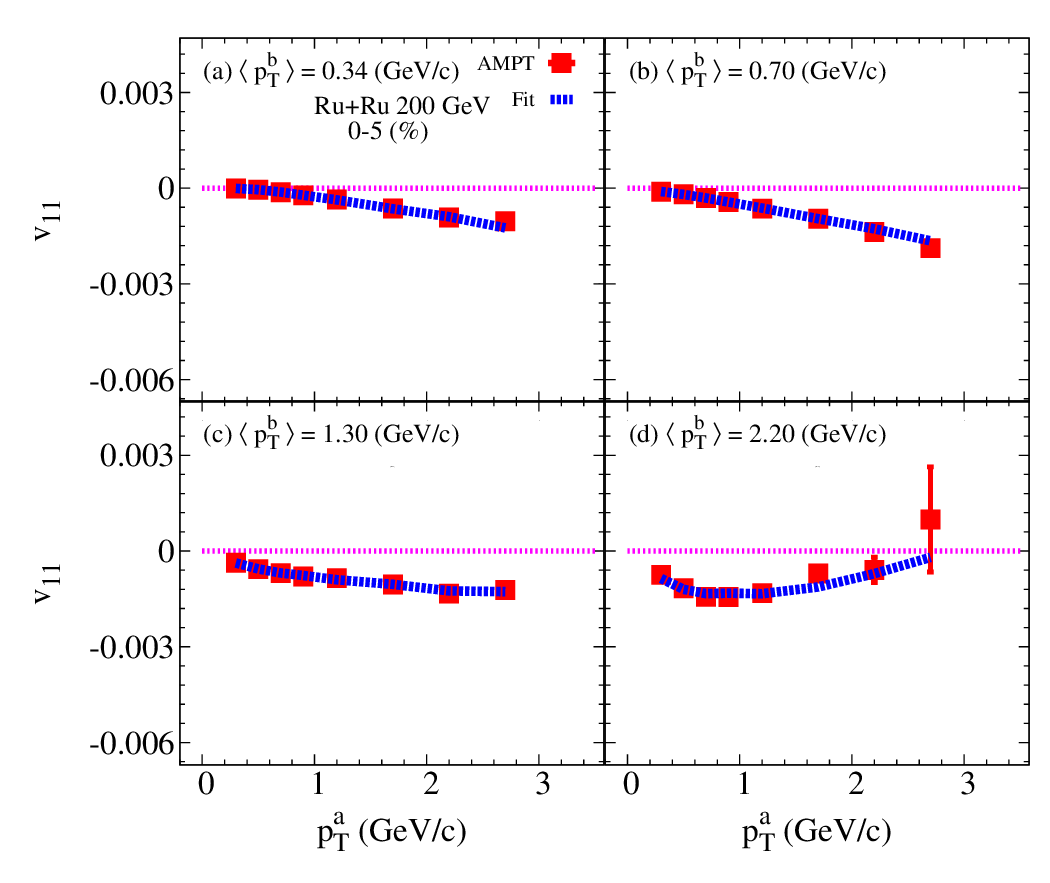}
\vskip -0.4cm
\caption{
$\mathrm{v_{11}}$ vs. $p_{T}^{\text{a}}$ for several $p_{T}^{\text{b}}$  for  0-5\% central Ru+Ru collisions [Case-1]. The curves indicate the result of the simultaneous fit with Eq.~(\ref{corrv1}).
}\label{fig:v11_pt}
\vskip -0.3cm
\end{figure}

For a single centrality selection, the $\mathrm{v_{11}}$ matrix is $\text{N}$-by-$\text{M}$ in size, and we fit it using the right-hand side of Eq.~\ref{corrv1}. The fit function encompasses $\text{N} + 1$ parameters: $N$ parameters correspond to $\mathrm{v^{even}_{1}(\pT)}$, and the parameter $\mathrm{K}$ accounts for momentum conservation~\cite{Jia:2012gu}. Figure~\ref{fig:v11_pt} illustrates the efficacy of the fitting approach for 0-5\% central Ru+Ru collisions. The curves in each panel depict the fits, showcasing how the $\mathrm{v_{11}}$ points evolve from negative to positive values as the selection range for $p_{T}^{\text{b}}$ is increased. This trend for $\mathrm{v_{11}}$ serves as an important constraint for the fits.

\begin{figure}[t]
\centering{
\includegraphics[width=0.9   \linewidth, angle=-0,keepaspectratio=true,clip=true]{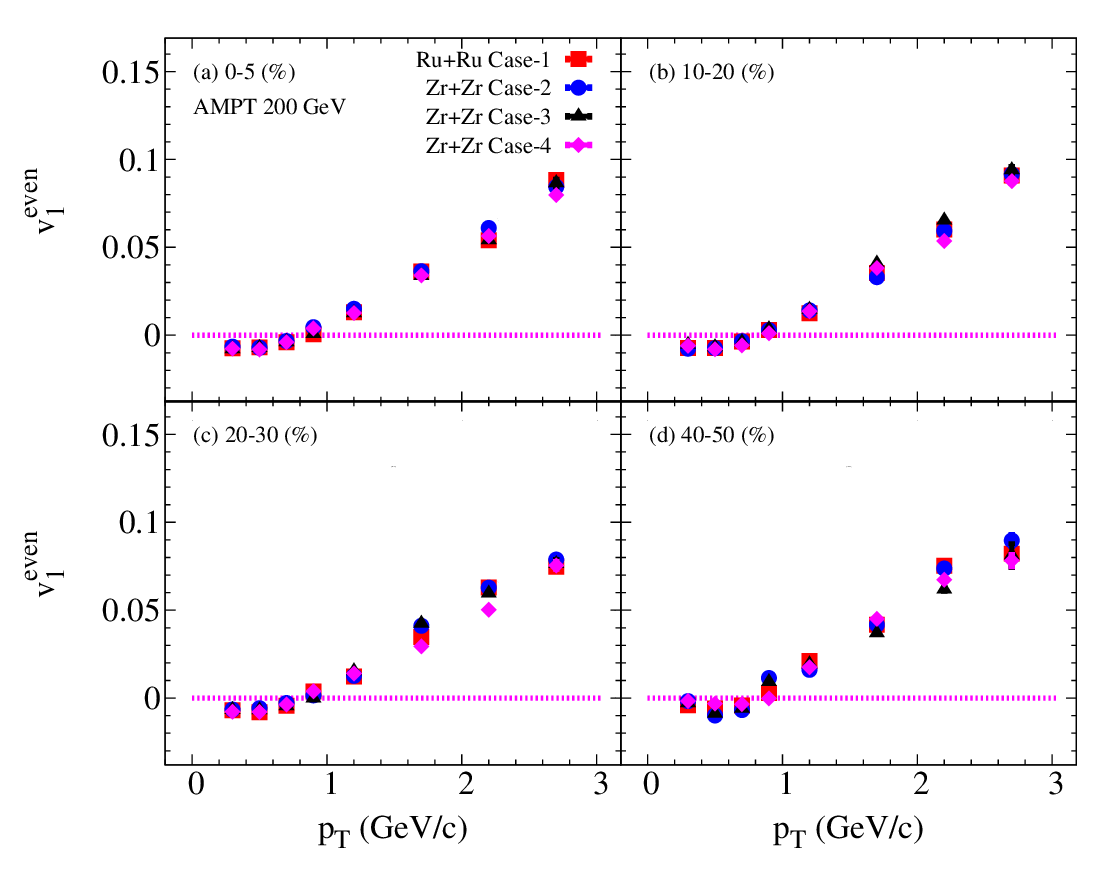}
 \vskip -0.3cm
\caption{ The extracted $\mathrm{v^{even}_{1}}$ vs. $p_T$ for 0-5\% (a), 10-20\% (b), 20-30\% (c), and 40-50\% (d)  \RuRu and \ZrZr collisions, for the respective cases in Table~\ref{tab:1} as indicated.
 }
  \label{fig:v1e_pt}
}
 \vskip -0.2cm
\end{figure}

The extracted ${v^{\rm even}_{1}}$ values for the respective cases in Table~\ref{tab:1} are depicted as functions of $p_T$ and centrality in Fig.~\ref{fig:v1e_pt}. They display a characteristic pattern~\cite{Retinskaya:2012ky}: a transition from negative $\mathrm{v^{even}_{1}(\pT)}$ at low $\pT$ to positive $\mathrm{v^{even}_{1}(\pT)}$ for $\pT \gtrsim 1$ GeV/c, with a crossing point that gradually shifts with increasing centrality. This behavior results from the requirement that the net transverse momentum of the system is zero, implying that the hydrodynamic flow direction of low $p_T$ particles opposes that of high $p_T$ particles. Additionally, the computations reveal small distinctions among the different AMPT Cases, which will be elaborated upon in subsequent discussions.
\begin{figure}[th]
\centering{
\includegraphics[width=0.9   \linewidth, angle=-0,keepaspectratio=true,clip=true]{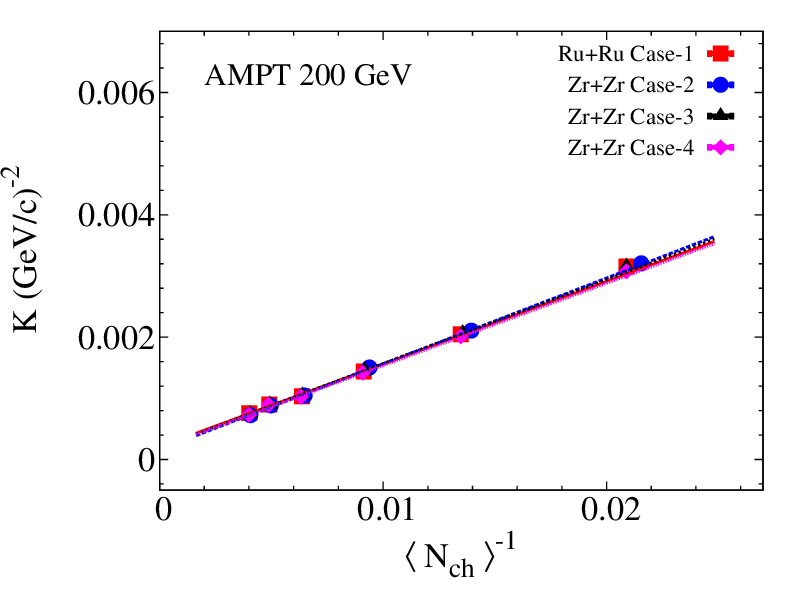}
\vskip -0.3cm
\caption{ $K$ values  vs. $\mathrm{\langle N_{ch} \rangle}^{-1}$ for the respective cases in Table~\ref{tab:1}. 
The lines show linear fits to the results.
 }
 \label{fig:K_pt}
}
\vskip -0.2cm
\end{figure}
The corresponding values of the GMC parameter $\mathrm{K}$ are shown in Fig~\ref{fig:K_pt}. These values reveal a linear correlation between $\mathrm{K}$ and $\mathrm{\langle N_{ch} \rangle^{-1}}$, mirroring the observed trend in experimental data~\cite{STAR:2019zaf}, albeit with a more modest slope in our model results. Furthermore, the slope of $\mathrm{K}$ with respect to $\mathrm{\langle N_{ch} \rangle^{-1}}$ is influenced by the nuclear structure parameters delineated in Table~\ref{tab:1}. This dependence indicates a subtle yet distinguishable disparity in slopes for Case-1 through Case-4, with a maximum difference of $\sim 4\% \pm 1.4$\% between Case-1 and Case-2. These insights imply that conducting meticulous comparisons of $\mathrm{K}$ vs. $\mathrm{\langle N_{ch} \rangle^{-1}}$ for data and theoretical models could offer a nuanced understanding of the GMC effect within theoretical frameworks. Furthermore, such analyzes could serve as a valuable tool for refining our grasp of the nuclear structure disparities between the Ru and Zr isobars.

\begin{figure}[th] 
\includegraphics[width=0.9  \linewidth, angle=-0,keepaspectratio=true,clip=true]{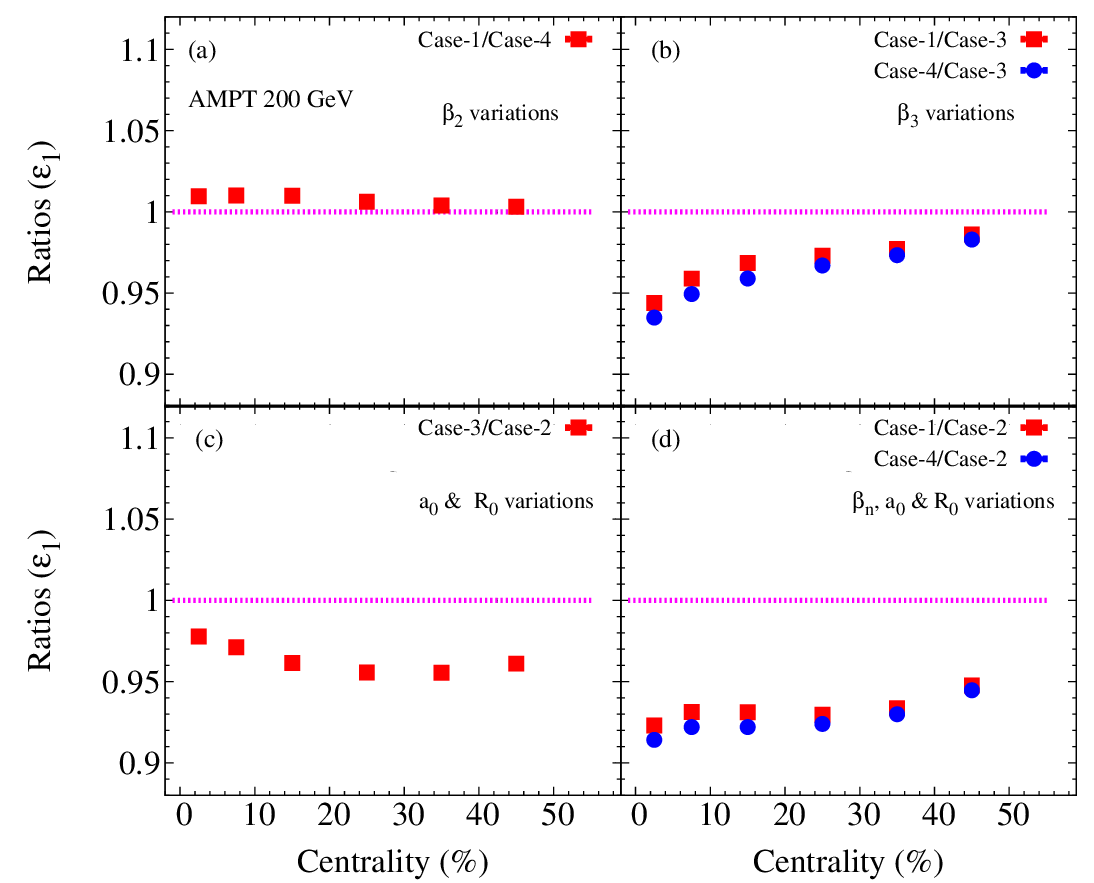}
\vskip -0.3cm
\caption{
The centrality dependence of the initial state $\epsilon_{1}$ ratios  for Case-1/Case-4 (a), Case-1/Case-3 and Case-4/Case-3 (b), Case-3/Case-2 (c), and Case-1/Case-2 and Case-4/Case-2 (d).
}\label{fig:e1_Cent_R}
\vskip -0.2cm
\end{figure}
To gain a deeper insight into the potential influence stemming from intrinsic deformation variations and neutron skin disparities between the isobaric nuclei, we investigated the ratios of $\varepsilon_1$ and $v^{\rm even}_{1}$ across the different scenarios outlined in Table~\ref{tab:1}. The centrality dependence of the $\varepsilon_1$ ratios are shown in Fig.~\ref{fig:e1_Cent_R}, with the intent of deciphering the interplay of distinct nuclear structure effects and capturing the sensitivity of $\epsilon_{1}$ to $\beta_n$, $R_0$, and $a_0$.

Analyzing the panels in Fig.~\ref{fig:e1_Cent_R}, we find that the Case-1/Case-4 ratio (panel a) reveals a resilience of $\epsilon_{1}$ against variations in $\beta_2$, as anticipated for a fluctuation-driven $\epsilon_{1}$. Contrarily, panels (b) show the Case-1/Case-3 and Case-4/Case-3 ratios, shedding light on a $\sim 3.0$\% difference in central collisions, which gradually diminishes as the collisions become more peripheral. This trend underscores the dependency of $\epsilon_{1}$ on $\beta_3$. Complementary insights into the effects of $a_{0}$ and $R_{0}$ on $\epsilon_{1}$ are gleaned from the Case-3/Case-2 ratios in panel (c), displaying an opposing trend compared to panel (b) with differences ranging from $\sim 1-3$\% based on centrality.

Combining the influences of $\beta_3$, $a_{0}$, and $R_{0}$ on $\epsilon_{1}$, panel (d) showcases the Case-1/Case-2 and Case-4/Case-2 ratios. These ratios indicate a $\sim 4.0$\% variation in central collisions that steadily diminishes as collisions become more peripheral. These findings collectively highlight the subtle but discernible sensitivity of the $\epsilon_{1}$ ratios to nuclear structure disparities between the two isobaric nuclei. This prompts the question: ''How does this sensitivity translate into the corresponding $\mathrm{v^{even}_{1}}$ ratios?''

\begin{figure}[th]
\centering{
\includegraphics[width=0.9 \linewidth, angle=-0,keepaspectratio=true,clip=true]{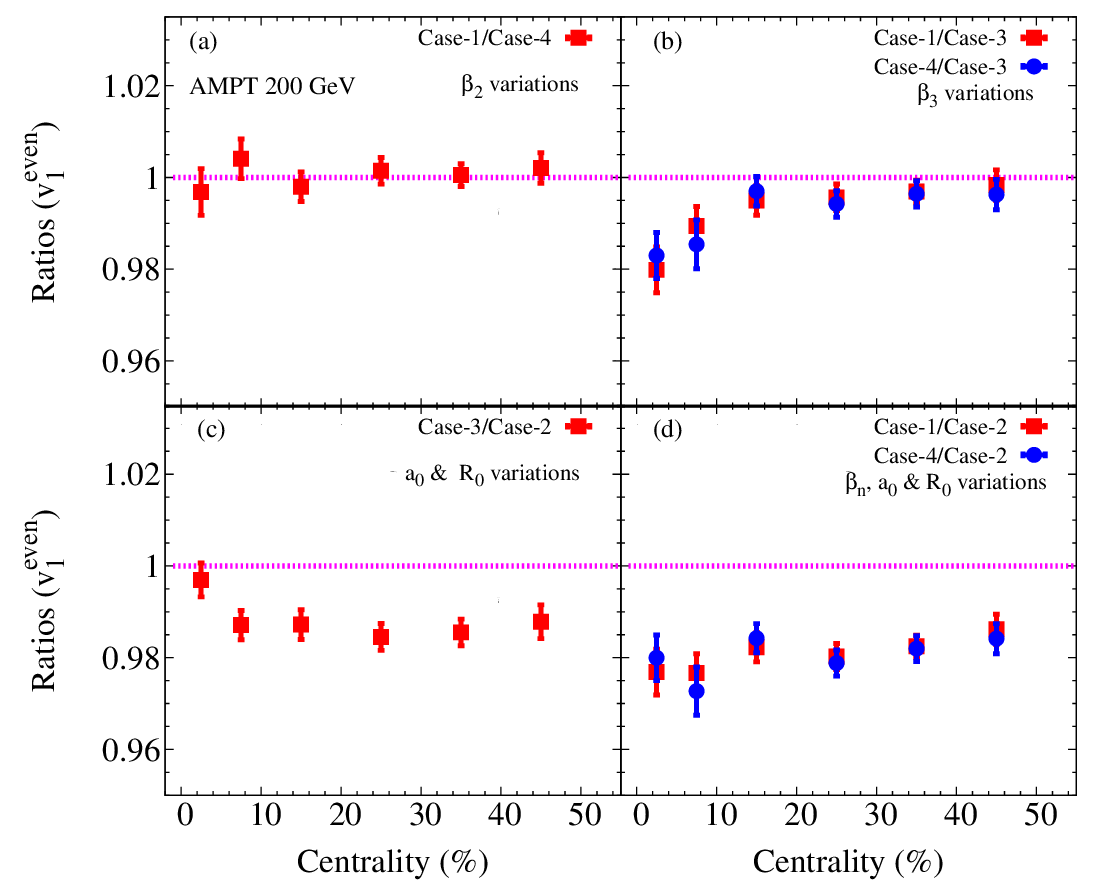}
\vskip -0.3cm
\caption{
Centrality dependence of the integrated ($0.2 < \pT < 0.9$~GeV/c) $\mathrm{v^{even}_{1}}$ ratios between Case-1/Case-4 (a), Case-1/Case-3 and Case-4/Case-3 (b), Case-3/Case-2 (c), and Case-1/Case-2 and Case-4/Case-2 (d).
}
\label{fig:v1_Cent_R}
}
\vskip -0.2cm
\end{figure}
The centrality-dependent $v^{even}_1$ ratios for the various AMPT cases are shown in Fig.~\ref{fig:v1_Cent_R} for the $0.2 < p{T} < 0.9$ GeV/c range. These ratios reflect patterns akin to those observed for the $\varepsilon_1$ ratios in Fig.~\ref{fig:e1_Cent_R}, albeit with reduced magnitudes. The Case-1/Case-4 ratio suggests the insensitivity of $v^{\rm even}_1$ to $\beta_2$ variations. In contrast, the Case-1/Case-3 and Case-4/Case-3 ratios hint at a $\sim 2$\% distinction in central collisions, which quickly diminishes towards peripheral collisions, showcasing the susceptibility of $v^{even}_1$ to $\beta_3$ variations. Panels (c) and (d) illuminate the subtle but perceptible response of $v^{\rm even}_1$ to variations in $a_0$ and $R_0$ and their combined impact. These distinctive trends suggests that $v^{\rm even}_1$ measurements for \RuRu and \ZrZr can serve as crucial constraints in delineating the nuclear structure differences between these isobaric nuclei.

\begin{figure}[th]
\centering{
\includegraphics[width=0.9  \linewidth, angle=-0,keepaspectratio=true,clip=true]{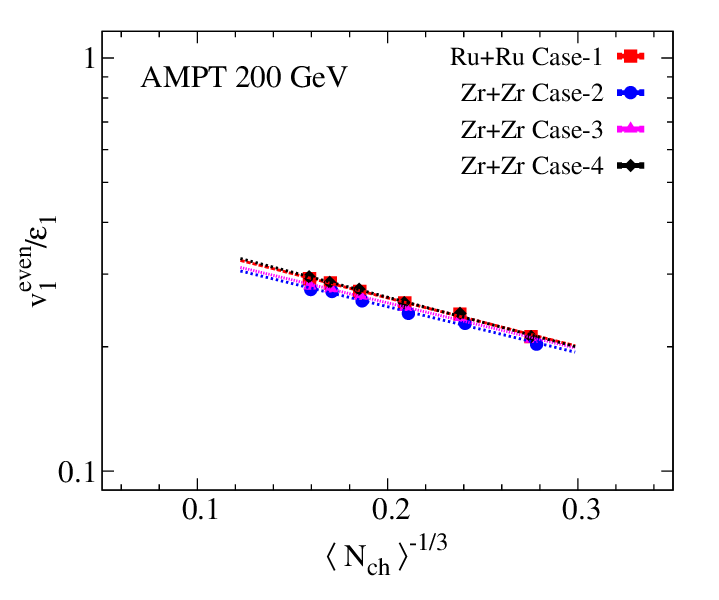}
\vskip -0.3cm
\caption{ Comparison of $v^{\rm even}_{1}/\varepsilon_1$ vs. $\mathrm{\langle N_{ch} \rangle}^{-1/3}$ for the respective cases in Table~\ref{tab:1} as indicated. The lines show the fits to the results.
 }
 \label{fig:E1_V1_Plot}
}
\vskip -0.1cm
\end{figure}
The magnitude of the slope in $\ln(v^{\rm even}_1/\varepsilon_1)$ vs. $\mathrm{\langle N_{ch} \rangle}^{-1/3}$ provides valuable insights into the viscous coefficient of the generated medium~\cite{Liu:2018xae, STAR:2019zaf}. As the effective viscous coefficient remains consistent across the scenarios outlined in Table~\ref{tab:1}, one can reasonably expect similar slope magnitudes in these simulated outcomes. As shown in Fig.~\ref{fig:E1_V1_Plot}, our results indeed exhibit this consistency, implying that comparing $\ln(v^{\rm even}{1}/\varepsilon_1)$ vs. $\mathrm{\langle N{ch} \rangle}^{-1/3}$ between \RuRu and \ZrZr for experimental data and $\varepsilon_1$ for the different cases could serve as an additional avenue to further refine the distinction of nuclear structure between these isobaric nuclei.

{\color{black}
Our investigation revealed a maximum deviation of 2\% and 4\% between \RuRu and \ZrZr for $\mathrm{v^{even}_{1}}$ and the parameter $K$, respectively. This observation prompts the question of whether such differences can be feasibly detected experimentally. In 2018, the STAR experiment at RHIC collected 1.2 billion minimum-bias events for each collision system (\RuRu and \ZrZr). These comprehensive data sets were gathered to (i) minimize statistical uncertainties in individual system measurements and (ii) mitigate systematic uncertainties in the \RuRu and \ZrZr ratios~\cite{STAR:2021mii}. Moreover, analysis conducted by the STAR experiment indicated that differences between the two systems could be discerned at levels approaching fractions of a percent in anisotropic flow measurements. Consequently, we advocate for the implementation of our proposed study by the STAR experiment at RHIC.
}

In summary, we have employed the two-particle correlation functions method within the AMPT model framework to examine the responsiveness of the rapidity-even dipolar flow ($\mathrm{v^{even}_{1}}$) and its associated Global Momentum Conservation parameter $K$ to inherent structural distinctions within $^{96}$Ru and $^{96}$Zr nuclei. Our model-based investigations have revealed modest yet discernible sensitivity to nuclear structure variations between the two isobars. This underscores the potential significance of measuring $\mathrm{v^{even}_{1}}$ and the GMC parameter in isobaric collisions at RHIC, offering a practical means to refine the theoretical model inputs essential for comprehending the nuclear structure disparities between the $^{96}$Ru and $^{96}$Zr isobars.

\section*{Acknowledgments}
The authors thank Dr.~Emily~Racow for the valuable discussions.
This research is supported by the US Department of Energy, Office of Nuclear Physics (DOE NP),  under contracts DE-FG02-87ER40331.A008.

\bibliography{ref} 
\end{document}